\documentclass[12pt,onepage,oneside]{article}
\usepackage{geometry}                
\usepackage{graphicx}
\usepackage{epstopdf}
\usepackage{etex}
\usepackage{tocloft}
\usepackage{booktabs}
\usepackage{natbib}
\usepackage{fancyhdr}
\usepackage{hyperref}
\usepackage{amsthm}
\usepackage{amsmath}
\usepackage{cleveref}
\usepackage{bbm}
\usepackage{multirow}
\usepackage{lscape}
\usepackage{bbm}
\usepackage{bm}
\usepackage{color}
\usepackage{setspace,caption}

\usepackage{ifthen}
\usepackage[latin1]{inputenc}
\usepackage{titletoc}
\usepackage{eso-pic}
\usepackage{float}

\usepackage{amssymb}
\usepackage{xcolor}

\def \E {\mathbb  E} 

\def \vol {v}

\usepackage{color,soul}
\definecolor{lightblue}{rgb}{.80,.95,1}
\sethlcolor{lightblue}
\setulcolor{red}

\title{Predicting the volatility of major energy commodity prices: the dynamic persistence model\thanks{We are grateful to the editor, Lance Bachmeier, and two anonymous referees for their useful comments and suggestions, which have greatly improved the paper. We are indebted to Rainer von Sachs, Federico Severino, Roman Liesenfeld, Nikolas Hautsch, Christian Hafner, Wolfgang H\"{a}rdle, Lubos Hanus for invaluable discussions and comments. We acknowledge insightful comments from numerous seminar presentations, such as: the Recent Advances in Econometrics (2023, Louvain), the 2021 and 2022 STAT of ML conference in Prague; the 15${\text{th}}$ International Conference on Computational and Financial Econometrics. The support from the Czech Science Foundation under the 19-28231X (EXPRO) project is gratefully acknowledged. For estimation of the quantities proposed, we provide a package \texttt{tvPersistence.jl} in \textsf{JULIA}. The package is available at \url{https://github.com/barunik/tvPersistence.jl}}}

\author{%
Jozef {\sc Barun\'{i}k}$^{\rm a,b}$\thanks{Corresponding author, Tel. +420 (776) 259 273, Email address: barunik@fsv.cuni.cz}, and
Luk\'{a}\v{s} {\sc V\'{a}cha}$^{\rm b,a}$
\vspace{5mm} \\
 \small $^{\rm a}$ Institute of Economic Studies, Charles University, \vspace{-0.5mm}\\ 
 \small $^{\rm b}$ The Czech Academy of Sciences, Institute of Information Theory and Automation}

\begin{document}
\maketitle

\begin{abstract}
\noindent Time variation and persistence are crucial properties of volatility that are often studied separately in energy volatility forecasting models. Here, we propose a novel approach that allows shocks with heterogeneous persistence to vary smoothly over time, and thus model the two together. We argue that this is important because such dynamics arise naturally from the dynamic nature of shocks in energy commodities. We identify such dynamics from the data using localised regressions and build a model that significantly improves volatility forecasts. Such forecasting models, based on a rich persistence structure that varies smoothly over time, outperform state-of-the-art benchmark models and are particularly useful for forecasting over longer horizons. \\

\noindent \textbf{Keywords}: persistence heterogeneity, wold decomposition, local stationarity, time-varying parameters \\
\noindent \textbf{JEL}:  C14, C18, C22, C50
\end{abstract}
\maketitle


\newpage 
\doublespace
\section{Introduction}

Energy commodities are of paramount interest to global economic prosperity because they are the most widely used source of energy.\footnote{To illustrate, more than 30\% of US energy consumption used petroleum-based fuels and another 30\% used natural gas. Most of this energy was used in transportation, industry and electricity generation. These figures are based on reports from the Lawrence Livermore National Laboratory: \url{https://www.llnl.gov}} They are constantly changing due to the constant influx of new technologies, environmental pressures and the geopolitical importance of controlling oil supplies. In recent years, large fluctuations in energy prices have been a major concern for both market participants and regulators, as price uncertainty has a significant impact on the economy \citep{elder2010oil} and measuring and predicting the variability of energy prices is essential for pricing derivatives, asset allocation or risk management, but is also a key factor in understanding fluctuations in stock prices, growth rates, inflation, employment and exchange rates \citep{hamilton1983oil,kilian2009not,kang2013modeling}. While an increasing number of authors pay considerable attention to forecasting volatility \citep{haugom2014forecasting,sevi2014forecasting,LU2022102218,zhang2019forecasting}, all approaches are challenged by severe limitations imposed by model assumptions. Determining the true data-generating process of volatility dynamics becomes a challenging and open question of great priority to market participants, financial analysts and policymakers.

In particular, it is well documented that energy volatility has exhibited a very high degree of temporal variation over the past decades, as both stable and uncertain periods have been driven by different shocks \citep{le2023volatility}. These can be associated with different states of an economy, as well as supply-side shocks, endogenous shocks driven by inventory and short-term price changes, or shocks due to the financialisation of commodities (and portfolio rebalancing). At the same time, an increasing number of authors have recently argued that a number of economic variables are driven by shocks that influence their future value with heterogeneous levels of persistence \citep{bandi2022spectral}. A possibly non-linear combination of transitory and persistent responses to shocks will produce volatility with heterogeneous persistence structures that remain hidden to the observer using traditional methods. In turn, research on energy volatility is limited to models that aggregate one of these features separately.

Inferring such time-varying persistence from data on energy prices and various measures of uncertainty has crucial implications for policy making, modelling or forecasting. However, despite progress in exploring structural breaks \citep{wen2016forecasting,arouri2012forecasting}, regime switching models \citep{ma2017forecasting}, popular heterogeneous autoregressive models \citep{zhang2019forecasting} or more complicated long memory structures \citep{wang2012long,wang2016forecasting, ozdemir2013persistence,charfeddine2014true,herrera2018forecasting}, which can exhibit large amounts of time persistence without being non-stationary \citep{baillie1996analysing}, there is still no clear consensus on how to explore such dynamic nature of the data. The inability to identify the dependence from the data alone leads to a tendency to rely on assumptions that are difficult, if not impossible, to validate. To better understand and forecast energy time series, we need an approach that can precisely locate the horizons and time periods in which the critical information occurs.

This paper proposes a novel representation for the non-stationary volatility of energy commodities that allows a researcher to identify and explore its rich time-varying heterogeneous persistence structures. We aim to identify localised persistence of volatility in major energy commodities that will be useful for modelling and forecasting purposes. Our work is closely related to the recent work of \cite{barunik2023persistence}, who propose to localise the persistence structure in time series and open new avenues for modelling and forecasting. In this paper, we develop such a model and open new avenues for modelling and forecasting in the energy economics literature.

Different degrees of persistence, which also change over time, are natural in energy commodity data. External shocks due to changing geopolitical risks and economic conditions affecting energy markets have different duration and persistence, depending on the behaviour and sentiment of agents.  The combination of shocks determines the persistence structure of energy commodity prices and, more importantly, their volatility. When analysing volatility time series using classical aggregate measures, we cannot see this complicated persistence structure with different shock durations \citep{ortu2020, bandi2021, barunik2023persistence}. Wold decomposition, the cornerstone of time series modelling that underlies the vast majority of models used, assumes that volatility is driven by a linear combination of shocks. For example, if the process has little dependence, then each shock will have a very short duration and we call this behaviour a transient. On the other hand, if the process is very persistent, close to a random walk, then a shock will remain in the time series for a long time. In a real time series we usually observe a mixture of transient and persistent effects of shocks. However, it is difficult to identify the different effects of shocks because the Wold decomposition aggregates the entire persistence structure. To see the full persistence structure, \citep{ortu2020} proposed the extended Wold decomposition, which allows to identify the effect of shocks at different horizons.

Our paper contributes to the literature exploring the persistence structure of volatility through time-varying changes in unconditional volatility, persistence and the occurrence of jumps \citep{granger1996varieties, granger2004occasional, bollerslev1993common, barunik2023persistence}. We argue that energy volatility has a rich time-varying persistence structure of shocks. To exploit these stochastic properties, we use a model that can capture the time variation of the persistence structure of volatility, namely the Time-Varying Extended Wold Decomposition (TV-EWD) model of \cite{barunik2023persistence}. To identify the time variation of the stochastic properties of volatility time series, the model uses the locally stationary process proposed by \cite{dahlhaus1996}. The localisation allows to obtain the locallised extended Wold decomposition, which allows a precise identification of the time-varying persistence structure. In the empirical part, we show that the time-varying persistence structure reveals significant changes in the way shocks propagate through the volatility of energy commodities. 

In the empirical part, we argue that the TV-EWD model is a useful tool for understanding and forecasting the volatility of energy commodities. Specifically, we use highly liquid crude oil, natural gas and regular gasoline futures contracts, which account for more than 75\% of the trading volume in the energy commodities market. Note that crude oil is the raw material used to produce heating oil, gasoline and other petroleum-based products. First, we identify the evolution of the time-varying persistence structure of volatility. Second, we build a model based on the identified structure and use it in a forecasting exercise. Our results indicate that forecasts based on the TV-EWD model are superior to other commonly used models, especially for longer horizon forecasts. As energy commodity prices and volatility are strongly related to the geopolitical situation, global economic conditions and agents' sentiment, it is natural to expect that the degree of persistence of shocks will vary over time. The empirical part discusses these persistence dynamics for events such as the annexation of Crimea in 2014, the Covid pandemic in 2020 and the aggression against Ukraine in February 2022. In contrast to these turbulent times, we also consider the calm period of 1990's when the volatilty was driven by very different structure of shocks.

The rest of the paper is structured as follows. Section 2 proposes a model for the time-varying persistence of energy commodities based on a locally stationary process, Section 3 examines the time-varying persistence of volatility in oil-based markets, and Section 4 concludes.


\section{Dynamically persistent volatility}
\label{sec:theory}

Here we present a model that captures the time-varying persistence structure of volatility. We begin by modifying the key idea of classical time series, which associates any covariance stationary time series with a linear combination of its own past shocks and moving average components of finite order \citep{wold1938, hamilton2020time}. While stationarity plays a key role in volatility analysis over decades due to the availability of natural linear Gaussian modelling frameworks, volatility turns out to be non-stationary in the longer run \citep{stuaricua2005nonstationarities}. The state of the world, as well as the behaviour of agents, is highly dynamic and the assumption of time-invariant mechanisms generating energy volatility is unrealistic. A more general non-stationary process can be one that is locally close to a stationary process at any point in time, but whose properties (covariances, parameters, etc.) gradually change in a non-specific way over time. The localisation of the classical representation is essential to allow time variation that captures unstable behaviour. Finally, we introduce the decomposition of shocks in a locally stationary volatility into a heterogeneous degree of persistence and propose a forecasting model that captures both the time variation and the persistence of shocks.

\subsection{Locally stationary volatility}

The idea that volatility can only be stationary for a limited period of time and that this is still valid for estimation was introduced by \cite{stuaricua2005nonstationarities}. Locally stationary processes that allow for slow variation of the stochastic properties of the process were formally introduced by \cite{dahlhaus1996}. 

Assume that a nonstationary volatility process $\vol_t$ depends on a time-varying parameter model. Following \cite{dahlhaus1996}, we replace $\vol_t$ by a triangular array of observations $(\vol_{t,T};t=1,\ldots,T)$ where $T$ denotes the sample size. We interpret the process $\vol_{t,T}$ as a local stationary approximation around a fixed point $t/T$. As a consequence, the process can change its stochastic properties smoothly over time. An important feature of this local approximation is the possibility to represent a locally stationary process $\vol_{t,T}$ as a time-varying MA($\infty$):

\begin{equation}
\label{eq:wold0}
\vol_{t,T} = \sum_{h=-\infty}^{+\infty} \alpha_{t,T} \left(h\right) \epsilon_{t-h},
\end{equation}
where the coefficients $\alpha_{t,T} \left(h\right) $ can be approximated under certain smoothness assumptions as $\alpha_{t,T} \left(h\right) \approx \alpha\left(t/T,h\right)$, see \cite{dahlhaus1996}. The innovations $\epsilon_t$ are independent random variables with $\E\epsilon_t = 0$, $\E|\epsilon_t| < \infty$ and $\E\epsilon_s\epsilon_t=0$ for $s\ne t$. The Wold decomposition of a locally stationary process in Eq.(\ref{eq:wold0}) is a linear combination of uncorrelated innovations with time-varying impulse response functions $\alpha\left(t,T\right)$ at a fixed point $t/T$.

\subsection{Time-Varying Extended Wold Decomposition (TV-EWD)}

Next, to identify the persistence structure of volatility, we use \cite{barunik2023persistence,ortu2020}, which decompose the (localised) time series into a collection of independent components. Each of these components represents a horizon (scale) with its own impulse response function that defines the degree of persistence at a given horizon and time. This allows us to decompose the responses of volatility to a unit shock into transient and persistent effects of these shocks in a time-varying manner. This variety of responses allows us to model the heterogeneity in the persistence structure that would otherwise remain hidden in an aggregate model.

In other words, we want to build a model in which the decomposed persistence structure changes smoothly over time. Such a Time-Varying Extended Wold Decomposition (TV-EWD) of volatility can significantly improve forecasting. The main assumption of the model is that we have a locally stationary (zero mean) process $\vol_{t,T}$, which has the representation $\vol_{t,T} = \sum_{h=-\infty}^{+\infty} \alpha_{t,T}(h) \epsilon_{t-h}$. Then, following Proposition 1 in \cite{barunik2023persistence}, for any $j \in \mathbb{N}, k \in \mathbb{N}$ we can define the decomposition of the persistence structure as:
\begin{equation}
\label{eq:tvewd}
\vol_{t,T}=\sum_{j=1}^{+\infty} \sum_{k=0}^{+\infty} \beta_{t,T}^{\{j\}}(k) \epsilon_{t-k2^j}^{\{j\}},
\end{equation}
where coefficients $\beta_{t,T}^{\{j\}}(k)= \frac{1}{\sqrt{2^j}}  \left[ \sum_{i=0}^{2^{j-1}-1} \alpha_{t,T}(k2^j+i) - \sum_{i=0}^{2^{j-1}-1} \alpha_{t,T}\left(k2^j+2^{j-1}+i\right) \right]$ denote the time-varying impulse response functions associated with scale $j$ and time-shift $k 2^j$ at a fixed point $t/T$, and $\sum_{k=0}^{\infty}\left(\beta_{t,T}^{\{j\}}(k)\right)^2 < \infty$ for all $j$. The innovations $\epsilon_t^{\{j\}} = \frac{1}{\sqrt{2^j}}  \left( \sum_{i=0}^{2^{j-1}-1} \epsilon_{t-i} - \sum_{i=0}^{2^{j-1}-1} \epsilon_{t-2^{j-1}-i} \right)$ are independent random variables with $\E\epsilon_t = 0$, $\E|\epsilon_t| < \infty$ and $\E\epsilon_s\epsilon_t=0$ for $s\ne t$. 

The TV-EWD model allows the time series to be decomposed into $j$ uncorrelated persistence components that can vary smoothly over time. Thus, it is a stationary approximation of the process $\vol_{t,T}$ with time-varying uncorrelated persistent components $\vol_{t,T}^{\{j\}}$:
\begin{equation}
\vol_{t,T}=\sum_{j=1}^{+\infty} \vol_{t,T}^{\{j\}}.
\end{equation}
Each component at scale $j$ can be written as: $\vol_{t,T}^{\{j\}}=\sum_{k=0}^{+\infty} \beta_{t,T}^{\{j\}}(k) \epsilon_{t-k2^j}^{\{j\}}$.  In other words, the decomposition allows us to study the time-varying impulse responses at different scales $j$. The shape of the impulse response provides information about how shocks propagate at a given scale at a given time. For example, with daily data, the first scale, $j=1$, shows how a unit shock propagates in 2 days, for the second scale, $j=2$, in 4 days, and so on.


\subsection{Identification of the dynamic persistence in volatility}

Having obtained the locally persistent representation of the volatility time series, we next estimate the parameters. Specifically, we start with the time-varying autoregressive (TVP-AR) coefficients model to compute the localised Wold decomposition. 

The TVP-AR(p) model of locally stationary process $\vol_{t,T}$ is for every fixed point $t/T$ defined as:
\begin{equation}
\label{eq:tvar}
\vol_{t,T}=\phi_0\left(t/T\right)+\phi_1\left(t/T\right)\vol_{t-1,T}+\ldots+\phi_p \left(t/T\right)\vol_{t-p,T} + \epsilon_t.
\end{equation}
 To obtain the time-varying coefficient estimates $\widehat{\Phi}\left(t/T\right)=\left(\widehat{\phi}_1\left(t/T\right),\ldots,\widehat{\phi}_p\left(t/T\right)\right)'$ we center the locally stationary process such that $\widetilde{\vol}_{t,T} = \vol_{t,T}-\widehat{\phi}_0\left(t/T\right)$. This setting is robust against a possible time trend. The coefficient functions $\phi_i\left(t/T\right)$ are estimated by the local linear method. This nonparametric regression approach has several advantages such as efficiency, bias reduction, and adaptation of boundary effects. For more details see \cite{fan1996local, barunik2023persistence}. 

Having the time-varying coefficients, we can compute the local Wold decomposition (Eq. \ref{eq:wold0}) and then further proceed to decompose the time-varying persistence structure using the TV-EWD (Eq. \ref{eq:tvewd}). Since we have a finite number of $T$ observations, we limit the depth of the decomposition to a finite number of scales $J$.
\begin{equation}
\widehat{\vol}_{t,T}=\sum_{j=1}^{J} \widehat{\vol}_{t,T}^{\{j\}} + \pi_t^{\{J\}}=\sum_{j=1}^{J} \sum_{k=0}^{N-1} \widehat{\beta}_{t,T}^{\{j\}}(k) \widehat{\epsilon}_{t-k2^j}^{\{j\}} + \pi_{t,T}^{\{J\}},
\end{equation}
where the estimate of (scale specific impulse response coefficients) $\widehat{\beta}_{t,T}^{\{j\}}(k)$ is computed as $\widehat{\beta}_{t,T}^{\{j\}}(k)= \frac{1}{\sqrt{2^j}}  \left( \sum_{i=0}^{2^{j-1}-1} \widehat{\alpha}_{t,T}(k2^j+i) - \sum_{i=0}^{2^{j-1}-1} \widehat{\alpha}_{t,T}(k2^j+2^{j-1}+i) \right)$, the scale dependent innovation process is obtained as $\widehat{\epsilon}_{t-k2^j}^{\{j\}} = \frac{1}{\sqrt{2^j}}  \left( \sum_{i=0}^{2^{j-1}-1} \widehat{\epsilon}_{t-i} - \sum_{i=0}^{2^{j-1}-1} \widehat{\epsilon}_{t-2^{j-1}-i} \right)$ and $\pi_{t,T}^{\{J\}}$ is the residual component defined that is usually very small hence we do not take it into account in the estimation. For more details, see \cite{ortu2020}.

\subsection{Forecasting procedure (model)}

We want to find a $h$ step-ahead forecast of the energy time series $\vol_{T+h,T}$. We have the localised version of the original, possibly non-stationary, time series $\vol_{1,T},\ldots,\vol_{T,T}$ and estimated values of $\widehat{\phi}_0\left(t/T\right)$, $\widehat{\beta}_{t,T}^{\{j\}}$ and $\widehat{\epsilon}_{t}^{\{j\}}$ we can proceed with the forecasting procedure. Since the scale components $j\in 1,\ldots,J$ have different information values, it is important to determine the importance of each component at horizon $j$. We do this using weights $w^{\{j\}}$, so we can write:
\begin{equation}
\vol_{t,T}=\widehat{\phi}_0\left(t/T\right) + \sum_{j=1}^{J} w^{\{j\}}\widehat{\vol}_{t,T}^{\{j\}}+\eta_{t,T}.
\end{equation}

Following \cite{barunik2023persistence}, we construct the conditional $h$-step-ahead forecasts as a sum of the conditional expected trend forecast $\E_t[\widehat{\vol}_{T+h,T}^{\{0\}}]$ and the weighted sum of scale components\footnote{We follow \cite{ortu2020} to forecast $\E_t[\vol_{t+1,T}^{\{j\}}]$} as:

\begin{equation}
\E_t[\vol_{T+h,T}]=\E_t[\widehat{\vol}_{T+h,T}^{\{0\}}] + \sum_{j=1}^{J} \widehat{w}^{\{j\}}\E_t[\vol_{T+h,T}^{\{j\}}]
\end{equation}

\section{Persistence structure of volatility in major energy commodities}

In this section, we demonstrate the importance of identifying the persistence structure in the volatility time series of energy commodities. Namely, we use the daily realised volatilities of three major energy commodities: crude oil, natural gas and regular gasoline. From the high frequency, irregularly spaced data\footnote{
obtained from Tick Data, Inc., which uses data from Globex.}, we extract 5-minute intraday prices of futures contracts, using the last available price in the 5-minute window. Contracts are automatically rolled over to provide continuous price records. We compute realised variance as the sum of squared intraday returns \citep{andersen2003modeling} computed from log prices and annualise it to volatility as $100\times\sqrt{252\times RV_t}$. 

It should be noted that the character of high-frequency futures data has changed dramatically over the decades, with major changes in the trading system, the introduction of near continuous trading on the CME Globex(R) electronic trading platform in December 2006, and the collection of data with milisecond time stamps in July 2011. In addition, the energy data have experienced both calm and turbulent periods where we can expect very different shocks to drive the time series. We therefore examine the performance of our model over two different periods that reflect these changes in the data. 

The first period covers January 2010 to December 2022, when energy prices were influenced by several turbulent sub-periods, such as COVID-19, and when milisecond timestamp data were collected almost all day on the electronic trading platform, resulting in large volumes. In contrast, the second period, from January 1993 to December 1999, was a calm and stable period for energy data after the decades of high volatility of the Great Inflation, and high frequency data were only available during open floor trading hours with a second time stamp. Futures were only available to customers through a broker, resulting in much lower average daily volumes compared to the upcoming periods. 

\subsection{Energy price volatility over the period 2010-2022}

\begin{figure}[ht!]
            \begin{center}
                \includegraphics[width=0.48\textwidth]{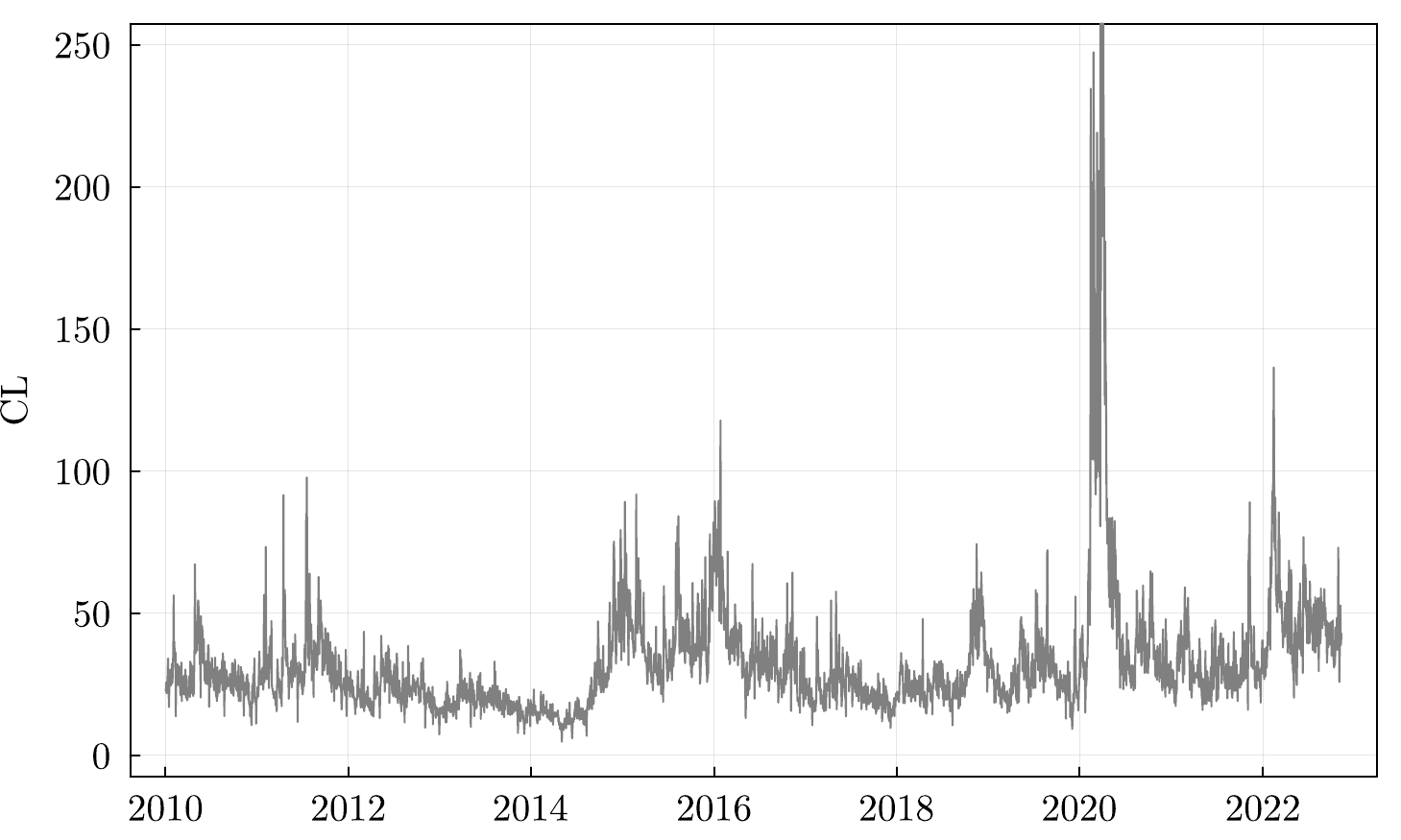}
                \includegraphics[width=0.48\textwidth]{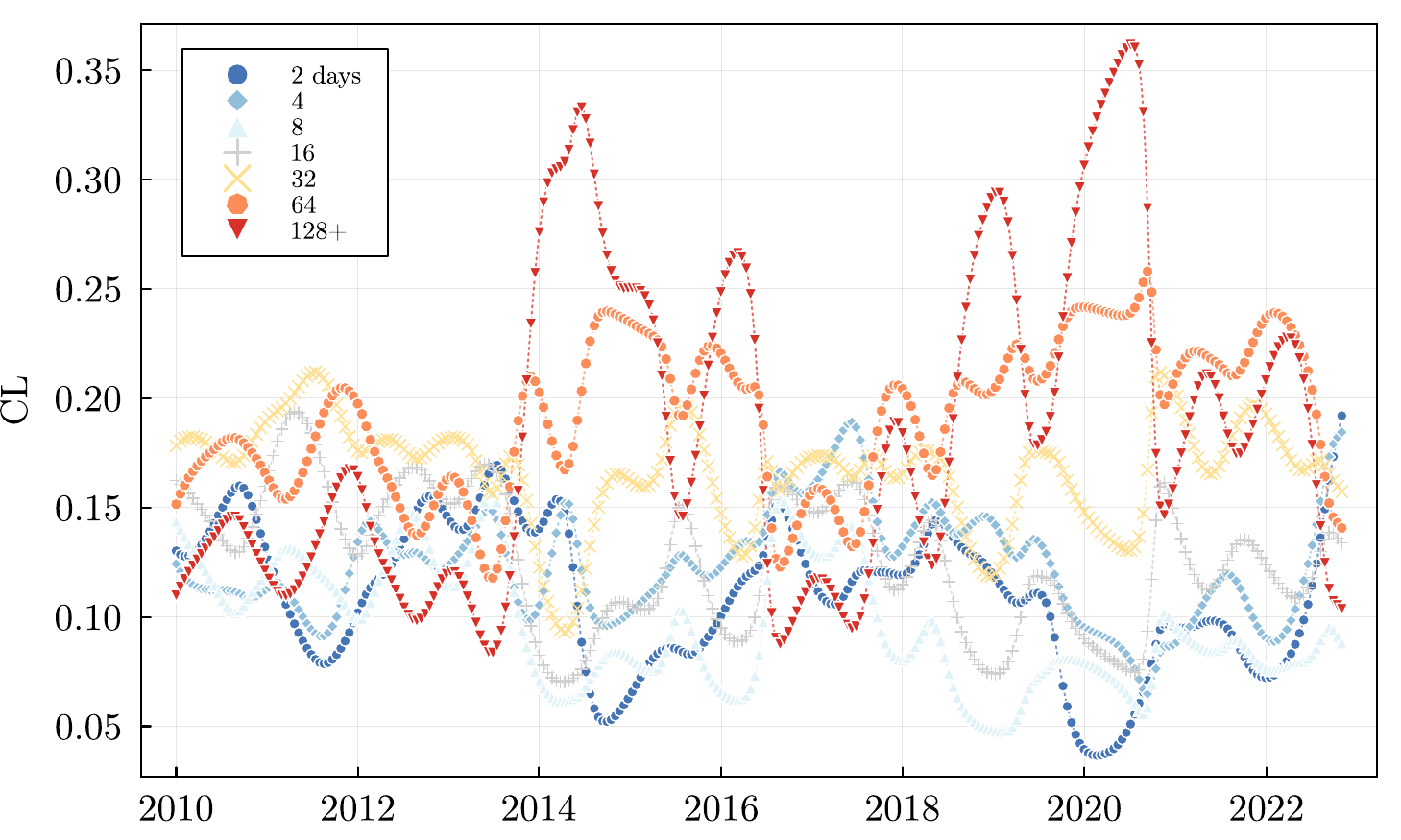} \\
                \includegraphics[width=0.48\textwidth]{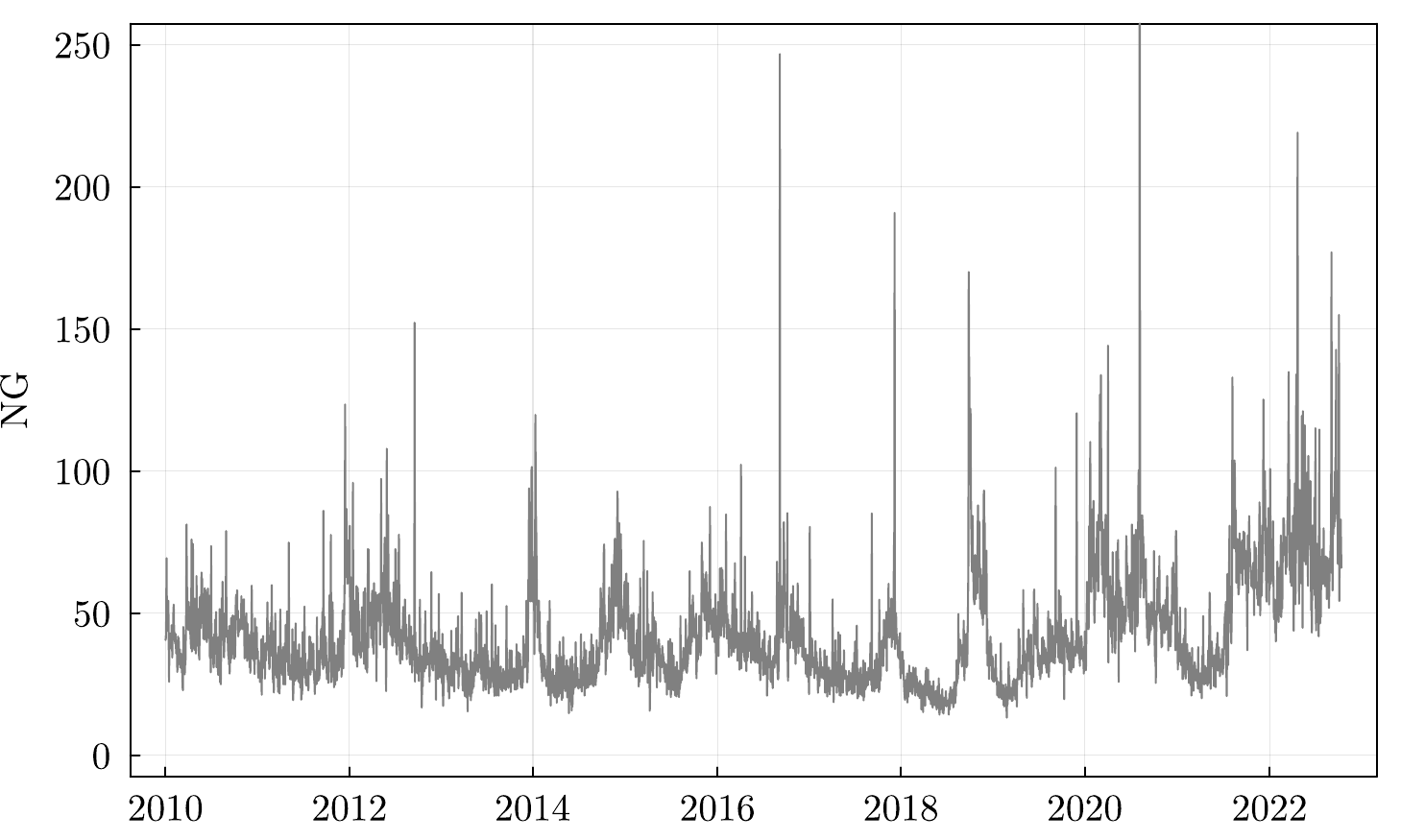}
                \includegraphics[width=0.48\textwidth]{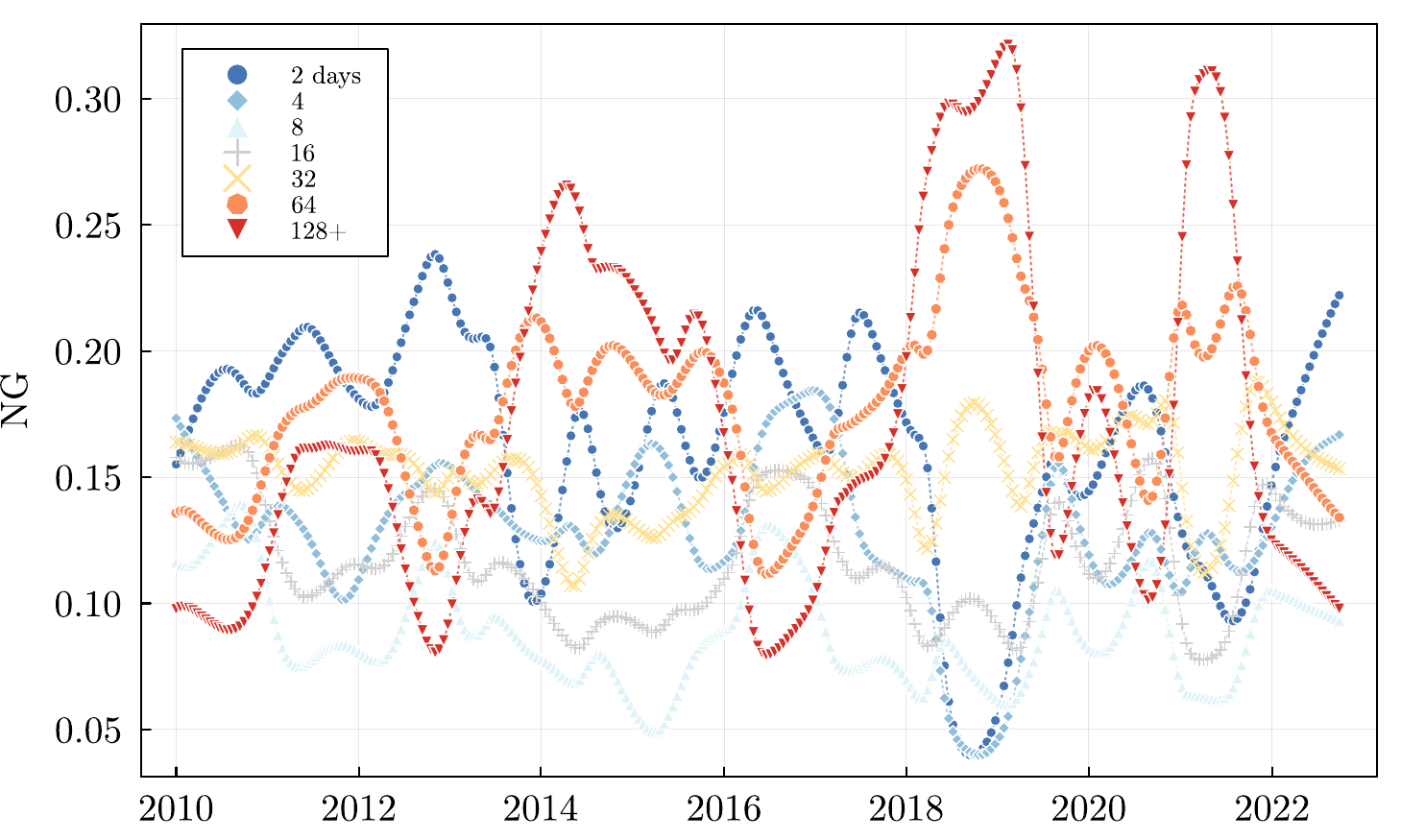} \\
                \includegraphics[width=0.48\textwidth]{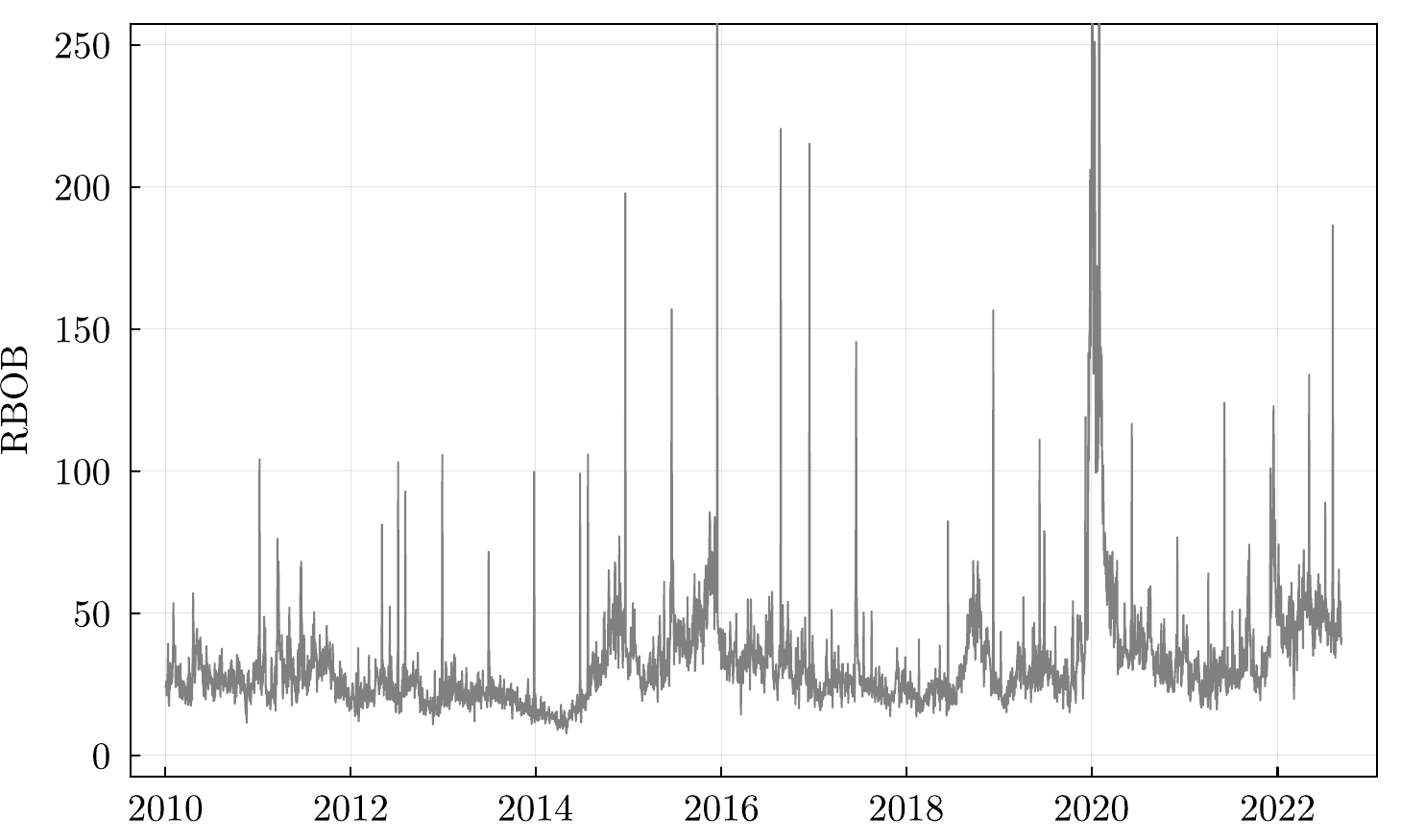}
                \includegraphics[width=0.48\textwidth]{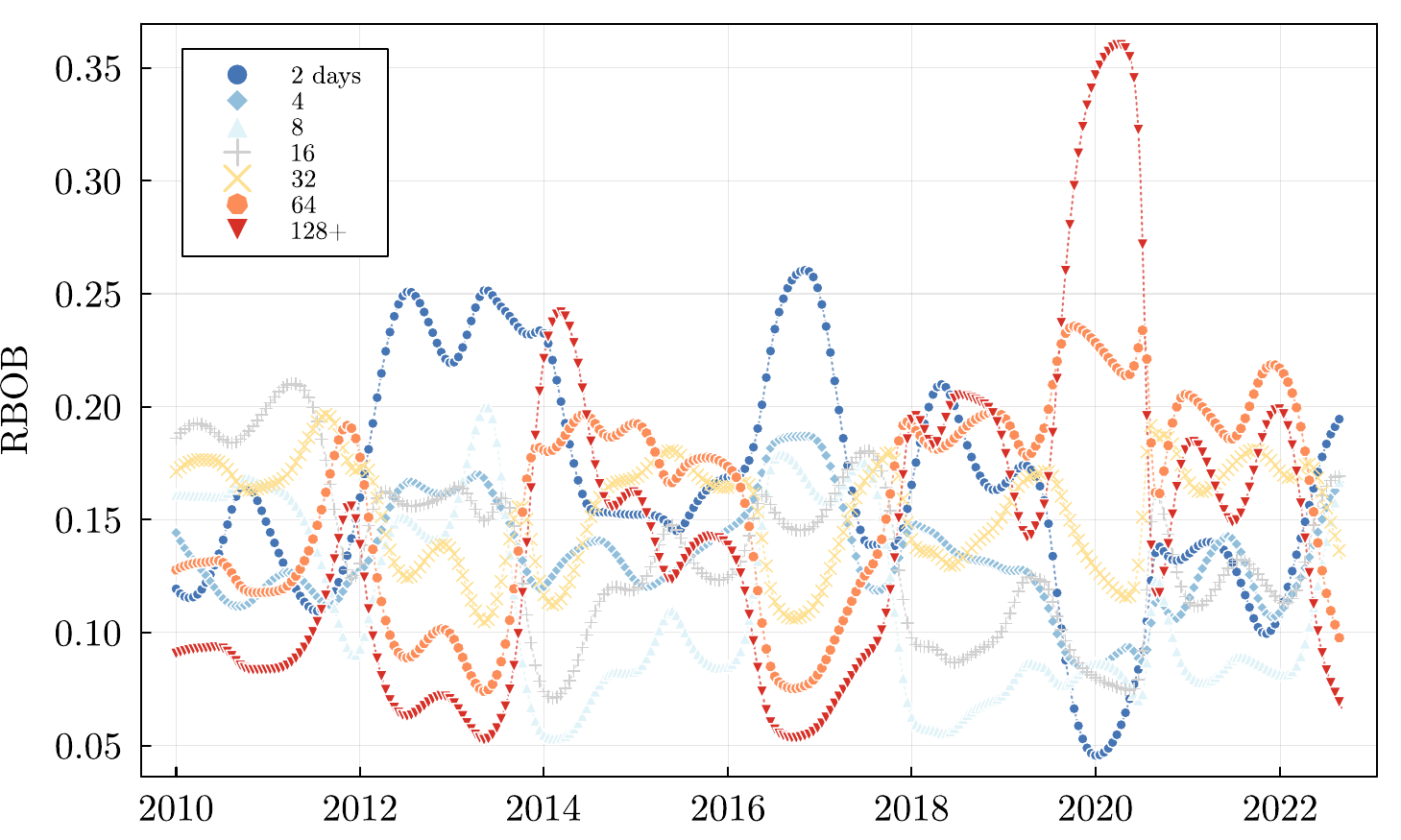}
            \end{center}
             \caption{\footnotesize{Left: Realized volatilities for crude oil (CL), natural gas (NG) and gasoline (RBOB). Right: Time-varying persistence structure of CL (top), NG (middle), RB (bottom). The plot shows the ratios of $\widehat{\beta}^{j}(t/T,1)/\sum_j \widehat{\beta}^{j}(t/T,1)$ on the y-axis, with $j$ corresponding to 2, 4, 8, 16, 32, 64, 128+ days persistence of the shocks represented by cold to warm colours, respectively, over the period from January 2010 to December 2022 on the x-axis.}}
        \label{fig:persistence_structure}
\end{figure}

During this period, futures contracts are traded on the New York Mercantile Exchange (NYMEX) and transactions are recorded with a milisecond timestamp on a 23-hour basis with a one-hour gap in trading. We redefine the day according to the electronic trading system. In addition, we exclude trades executed on US federal holidays, 24-26 December and 31 December to 2 January, due to low liquidity on these days, which could lead to an estimation bias. The data cover the period from January 2010 to December 2022. The left column of the figure (\ref{fig:persistence_structure}) shows the time series for all three commodities.

Following \cite{barunik2023persistence}, we use the multiscale impulse response functions from the TV-EWD model. Looking at the evolution of the volatility persistence structure gives us detailed information about the persistence of the shocks that generate the volatility time series of the energy commodities under study.

We begin by illustrating the persistence structure derived by our model. Figure (\ref{fig:persistence_structure}) shows the evolution of the persistence structure of volatility. We report the persistence at a given scale $j$ as the persistence share of that scale relative to the persistence at other scales. More specifically, we compute the ratio of the first coefficient of the multiscale impulse response function $\widehat{\beta}^{j}(t/T,1)$ to the sum of the first coefficients of all scales $\sum_j \widehat{\beta}^{j}(t/T,1)$. The scales $j$ represent 2,4,8,16,32,64,128 days persistence of shocks and are represented in the figure from cold (scale 2) to warm (scale 128) colours (\ref{fig:persistence_structure}). As the persistence structure is highly heterogeneous and time-varying, it motivates the use of the TV-EWD model for modelling and forecasting volatility. There are four notable sub-periods in which the persistence structure of volatility changes significantly:
\begin{itemize}
\item \textbf{Before 2014}: The volatility of CL is dominated by components calculated from shocks of medium persistence (16-64 days), with transitory shocks (2-8 days) of rather low importance. In contrast, NG and RBOB volatilities are dominated by transitory shocks.

\item \textbf{Feb 2014 (annexation of Crimea)}: The importance of the more persistent components of oil and gasoline has increased significantly, with gasoline spiking only for very short periods. In December 2014, the oil price was 50\% lower than in April 2014, and oil volatility is mostly driven by persistent shocks of 64-128 days, which generate about 60\% of the shocks in this period. Gas and petrol also show higher persistence.

\item \textbf{Onset of the Covid-19 pandemic}: The beginning of 2020 is characterised by the oil price spike followed by a sharp drop, leading to very high volatility. Interestingly, components driven by shocks with a persistence of 64 and 128+ days generate more than 50\% of the volatility. Gasoline shows similar behaviour to oil. However, natural gas does not react as strongly to the pandemic, as we observe a mixture of transient and persistent shocks (almost homogeneous structure with equal persistence). 

\item \textbf{Feb 2022}: After the start of the aggression in February 2022, we observe a strong dynamic with the dominance of persistent shocks being overtaken by transitory shocks (2 and 4 day persistence accounts for 40\%). For oil, transitory shocks also dominate in this period, but they are not as strong (2 and 4 day persistence accounts for 35\%). This is a consequence of a very unstable and rapidly changing situation, in which transitory shocks to market participants' expectations generated a large proportion of price fluctuations. After February 2022, we clearly observe an increase in overall volatility for all three commodities with mixed persistence structures.
\end{itemize}

Note that although the level of volatility in these periods is very different, especially Covid-19 is characterised by very high volatility, all turbulent sub-periods are driven by shocks with higher persistence due to the increased uncertainty of market participants. Therefore, compared to other calmer periods, we can observe a higher proportion of higher persistence components driving the series. We will return to this discussion in a later section that examines the calm period of the 1990s. It is also interesting to note that this is not the case for natural gas, which is driven by transitory components after February 2022. With the immense impact of the Russian invasion on the importance of gas, we observe large fluctuations in volatility after February 2022, showing that a lot of transitory uncertainty is driving the time series.

\subsubsection{Volatility forecasts}

Having identified a rich dynamic in the volatility persistence of the three major energy commodities, we use it to build the forecasting model. The time-varying persistence structure shows that we often witness changes from the transient to the persistent nature of the shocks that dominate volatility. This motivates us to use the TV-EWD model, which can capture the non-trivial, possibly non-stationary dynamics of the volatility series. 

To compare the forecasting performance of the TV-EWD model with other commonly used models for estimating the volatility of energy commodities, we use the models that best approximate persistence or long memory behaviour, such as the Heterogeneous Autoregressive Model (HAR) of \cite{corsi2009} and the EWD model of \cite{ortu2020}. This model also uses the extended Wold decomposition, but faces the same difficulties in capturing the time-varying nature of volatility time series.\footnote{In contrast to the paper by \cite{ortu2020}, we use a shorter in-sample period and depth of decomposition. This allows the model to partially account for changes in the persistence structure of volatility.} While both models capture the unconditional long-run dependence well, they assume stationarity of volatility. To assess the dynamic behaviour, we use both the simple TVP-AR(3) model and the TV-HAR model. The latter model is powerful in capturing the time-varying structure as well as in exploiting the heterogeneous horizon structure (persistence) as it allows for daily, weekly and monthly horizons. This makes the model very versatile and usually hard to beat as it is both highly effective and parsimonious. We save the first 700 observations for the in-sample fit and obtain 2600 forecasts from a rolling window starting the out-of-sample forecasting period with September 15, 2012.

Note that the kernel width is set to 0.3 and the decomposition depth to 7 scales, which is similar for all TV-EWD model settings. This means that the longest horizon considered is 128 days and more (128+). However, the TV-AR(p) model differs for both forecast horizons and commodities. The TV-AR(p) model for CL has $p=2$ for the 1-day-ahead forecast and $p=6$ for the 5- and 22-day-ahead forecasts. For NG the TV-AR(p) model has $p=5$ for all forecasts and for RB the TV-AR(p) model has $p=2$ for the 1-day-ahead forecast and $p=5$ for the 5- and 22-day-ahead forecasts. All parameters, such as the number of lags of the approximating model, the kernel widths or the degree of decomposition, have been chosen with a view to improving the out-of-sample performance for a given time series and thus reducing losses. One of the key parameters of the model is the number of lags $p$ in the TV-AR($p$), which determines the decomposition. For longer horizon forecasts, it is crucial to choose a larger $p$ as it helps to capture a richer structure in terms of persistence. On the other hand, for $h=1$ forecasts, a more parsimonious model with fewer lags helps to reduce the error. Again, this parameter is selected for each forecast horizon and time series based on the largest reduction in losses.

\begin{table}[ht!]
\footnotesize 
\caption{\footnotesize{The 2010-2022 period: Root mean square error (RMSE) and mean absolute error (MAE) of the TV-EWD model compared to \cite{ortu2020}'s extended wold decomposition (EWD), time-varying autoregression (TV-AR(3)). All errors are relative to the time-varying HAR (TV-HAR) model of \cite{corsi2009} over $h=1$, $h=5$ and $h=22$. Thus, the value bellow one implies that the corresponding model outperforms the TV-HAR benchmark. We use $*$,$**$ and $***$ to denote that a competing model has significantly lower MSE and MAE (is better) in comparison to the TV-HAR model at 90\%, 95\% and 99\% significance levels, respectively. We use $\dag$,$\dag \dag$ and $\dag \dag \dag$ to denote that a competing model has significantly higher MSE and MAE (is worse) in comparison to the TV-HAR model at 90\%, 95\% and 99\% significance levels, respectively.}}

\begin{center} 
\begin{tabular}{lllllllll} 
\toprule 
& & \multicolumn{3}{c}{RMSE} & & \multicolumn{3}{c}{MAE} \\ 
\cmidrule(r){3-5} \cmidrule(r){7-9}
& & $h=1$ & $h=5$ & $h=22$ & & $h=1$ & $h=5$ & $h=22$ \\ 
\cmidrule(r){3-5} \cmidrule(r){7-9}
& EWD &  1.373$^{\dag \dag \dag}$ & 1.542$^{\dag \dag \dag}$ & 0.962 & & 1.457$^{\dag \dag \dag}$ & 1.577$^{\dag \dag \dag}$ & 1.185$^{\dag \dag \dag}$ \\
CL & TV-AR3 & 1.051$^{\dag \dag}$ & 1.062$^{\dag}$ & 0.855$^{***}$ & & 1.035$^{\dag \dag \dag}$ & 1.108$^{\dag \dag \dag}$ & 1.024\\
& TV-EWD & 1.009 & 0.946$^{*}$ & 0.807$^{***}$ & & 1.008 & 0.982 & 0.960$^{**}$\\
\cmidrule(r){3-5} \cmidrule(r){7-9}
& EWD &   1.297$^{\dag \dag \dag}$ & 1.381$^{\dag \dag \dag}$ & 1.235$^{\dag \dag \dag}$ & & 1.437$^{\dag \dag \dag}$ & 1.621$^{\dag \dag \dag}$ & 1.358$^{\dag \dag \dag}$ \\
NG & TV-AR3 & 1.064$^{\dag \dag \dag}$ & 1.134$^{\dag \dag \dag}$ & 1.300$^{\dag \dag}$ & & 1.089$^{\dag \dag \dag}$ & 1.253$^{\dag \dag \dag}$ & 1.234$^{\dag \dag \dag}$\\
& TV-EWD & 1.003 & 0.945$^{**}$ & 0.923$^{***}$ & & 1.006 & 0.965$^{***}$ & 0.895$^{***}$\\
\cmidrule(r){3-5} \cmidrule(r){7-9}

& EWD & 1.249$^{\dag \dag \dag}$ & 1.444$^{\dag \dag \dag}$ & 1.125$^{\dag \dag \dag}$ & & 1.481$^{\dag \dag \dag}$ & 1.535$^{\dag \dag \dag}$ & 1.299$^{\dag \dag \dag}$\\
 RB & TV-AR3 & 1.052$^{\dag \dag \dag}$ &  1.112$^{\dag \dag \dag}$ & 0.997 & & 1.084$^{\dag \dag \dag}$ & 1.186$^{\dag \dag \dag}$ & 1.105$^{\dag \dag \dag}$\\
& TV-EWD & 1.000 & 0.954$^{*}$ & 0.880$^{***}$ & & 1.017 & 1.003 & 0.951$^{***}$\\
\bottomrule 
\end{tabular} 
\end{center} 
\label{tab:RVs} 
\end{table}

Table \ref{tab:RVs} summarises the results of the volatility forecasting performance of the main energy commodities. We report the Root Mean Square Error (RMSE) and the Mean Absolute Error (MAE) loss functions. The TV-HAR model is used as a benchmark and all reported losses are relative to the TV-HAR. Thus, a value less (greater) than one indicates that the corresponding model has a better (worse) forecasting performance relative to the benchmark.

In general, the ability of the TV-EWD model to outperform all alternatives increases with increasing forecast horizons. TV-EWD provides significantly better forecasts than the benchmark in all cases for forecasts of 22 days ahead. The superior performance in the MAE also suggests that the forecasts are not systematically biased relative to the other models. 

For the 5-day-ahead forecasts, the results are mixed. In terms of RMSE, the TV-EWD again outperforms the benchmark significantly. In terms of MAE, our model still provides significantly better forecasts for natural gas, but the forecasts for the other two commodities are indistinguishable from the TV-HAR. Similarly, the forecast errors of our approach and TV-HAR are the same for the 1-day-ahead forecast for both the RMSE and MAE loss functions.

The EWD model of \cite{ortu2020} performs well for the 22-day-ahead forecasts as measured by the RMSE loss. This implies that the model is immune to large forecast outliers. Conversely, the poor performance, as measured by the MAE, indicates a systematic bias in the forecast. This is caused by a very slow adjustment to changes in the level of long-term volatility. This is a consequence of the model's lack of time-varying capabilities.

\subsection{Energy price volatility of the 1993-1999 period}

\begin{figure}[ht!]
            \begin{center}
                \includegraphics[width=0.48\textwidth]{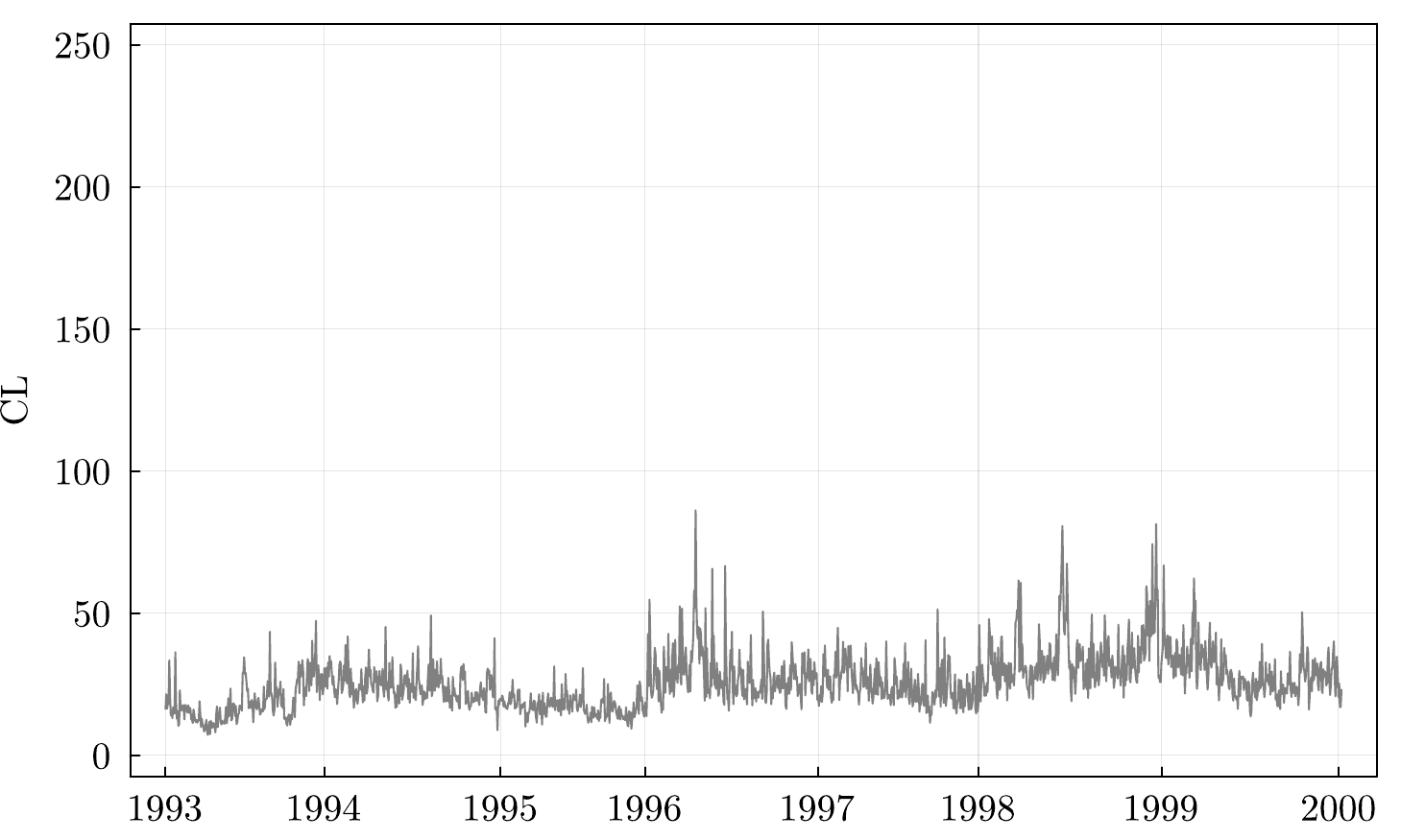}
                \includegraphics[width=0.48\textwidth]{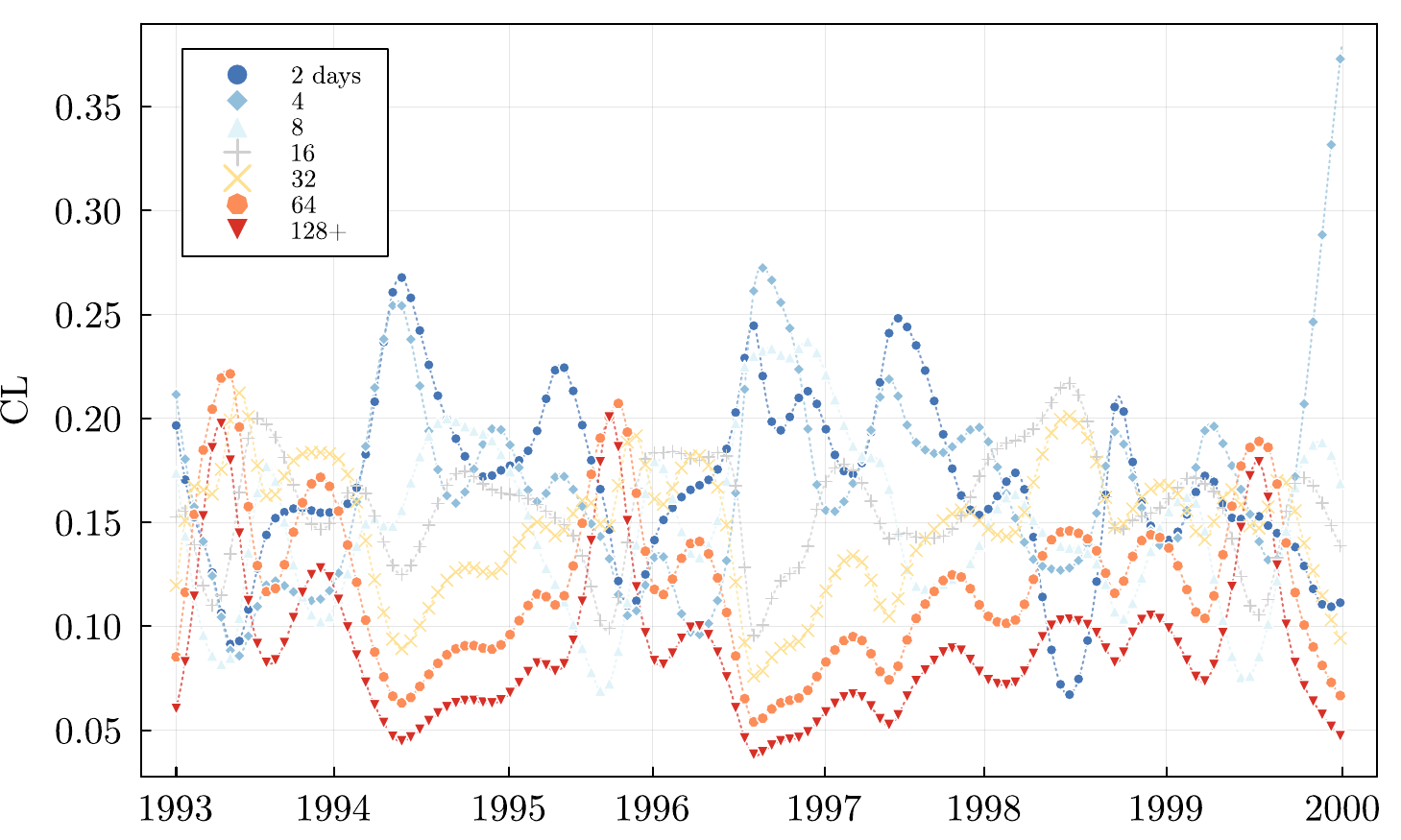} \\
                \includegraphics[width=0.48\textwidth]{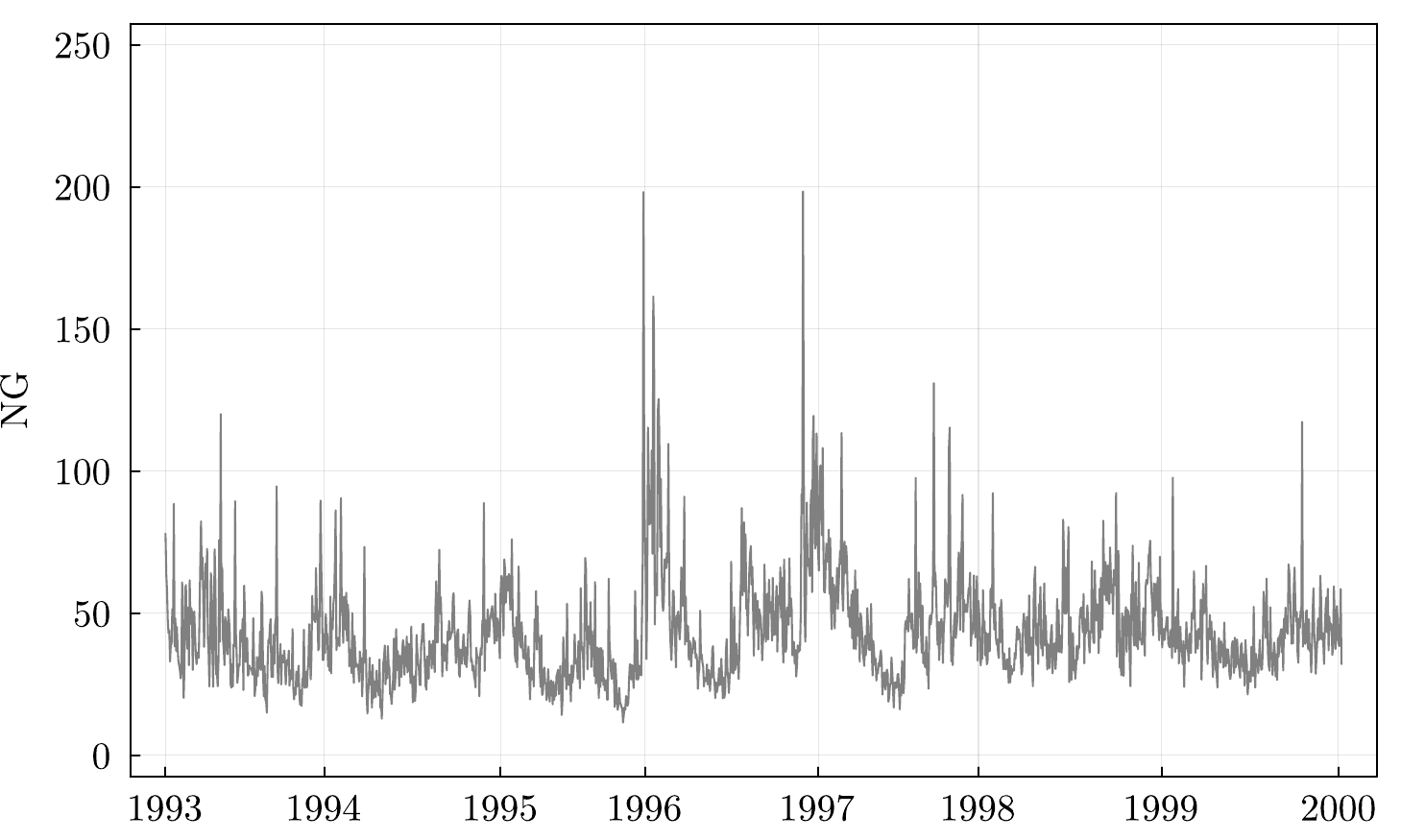}
                \includegraphics[width=0.48\textwidth]{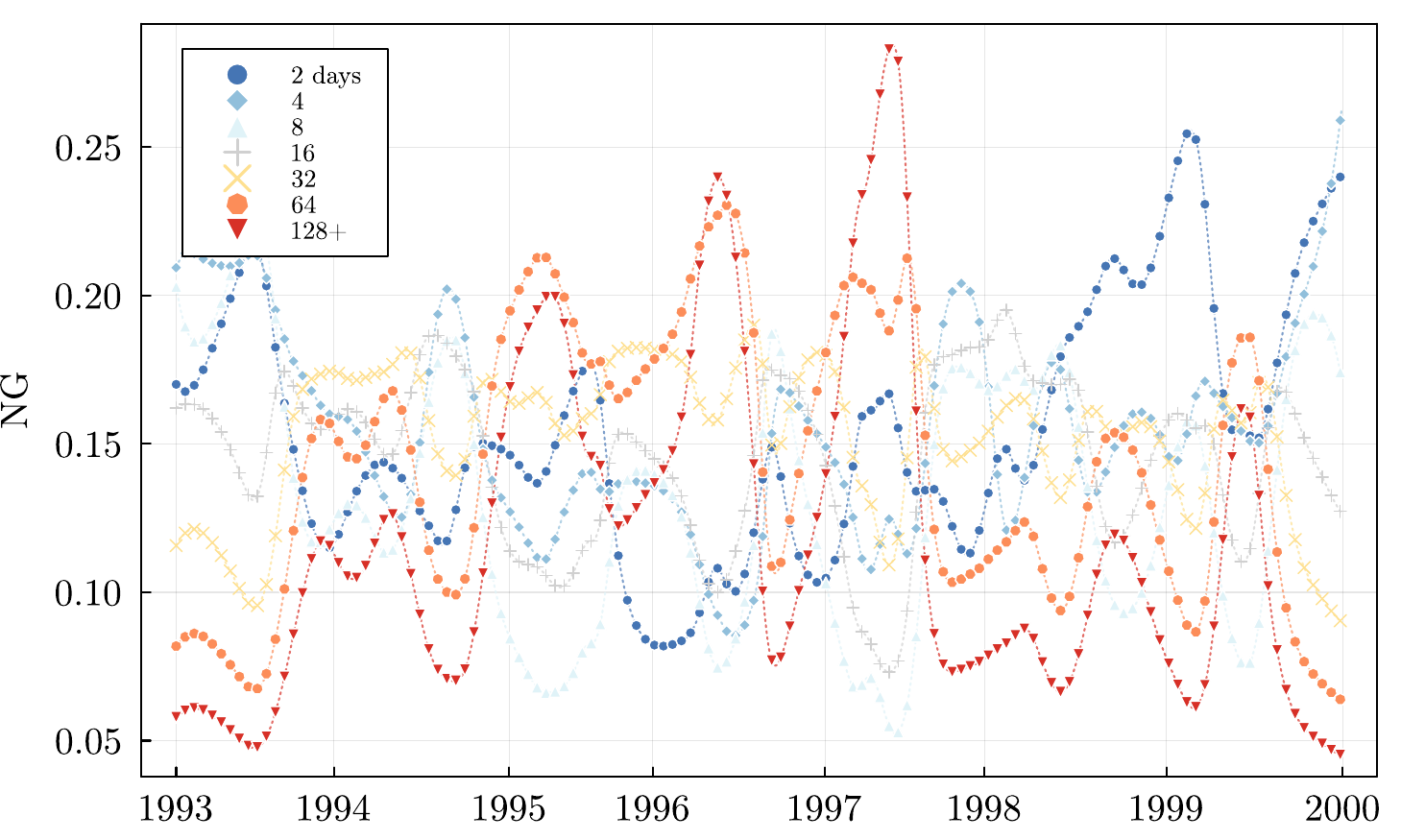} \\
                \includegraphics[width=0.48\textwidth]{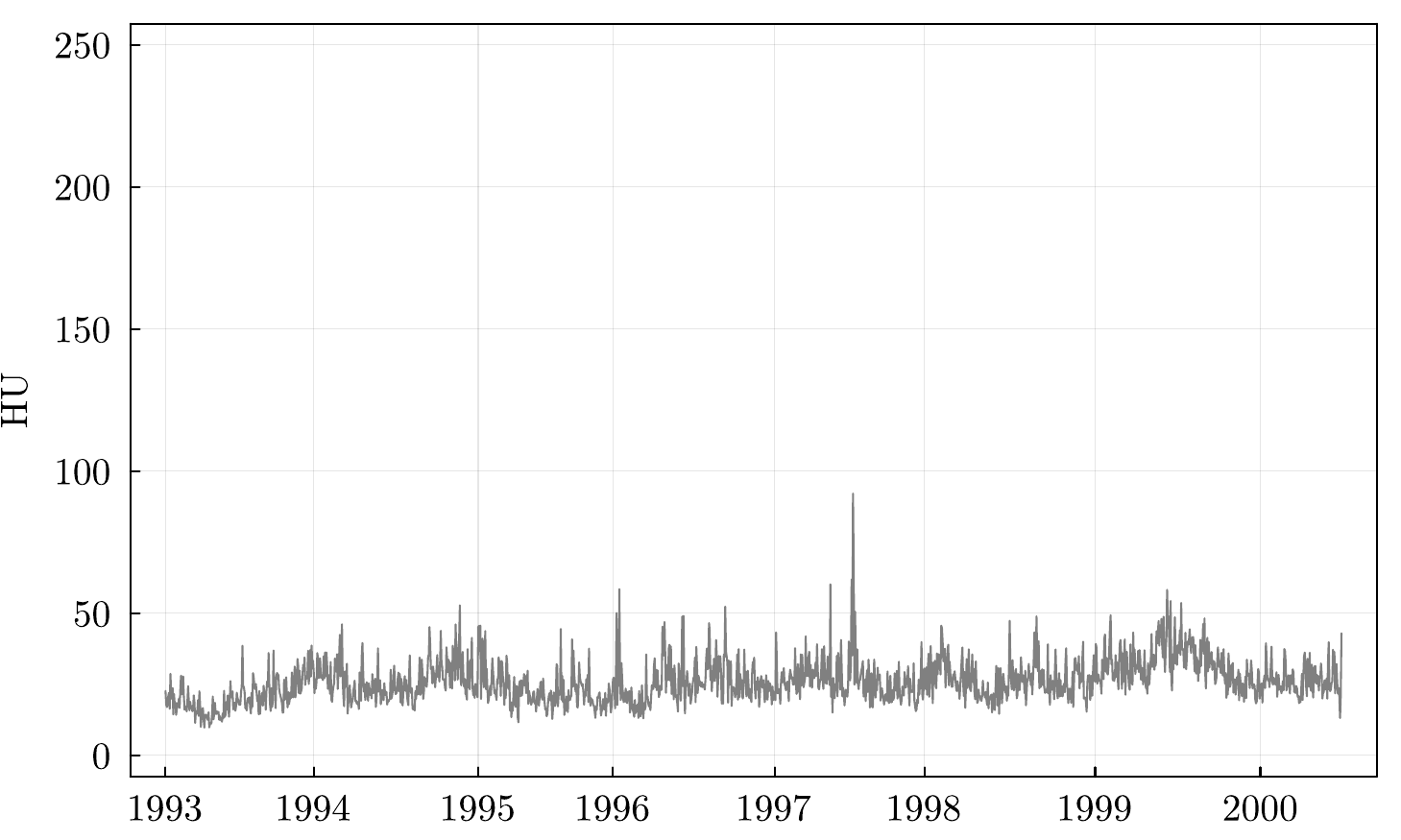}
                \includegraphics[width=0.48\textwidth]{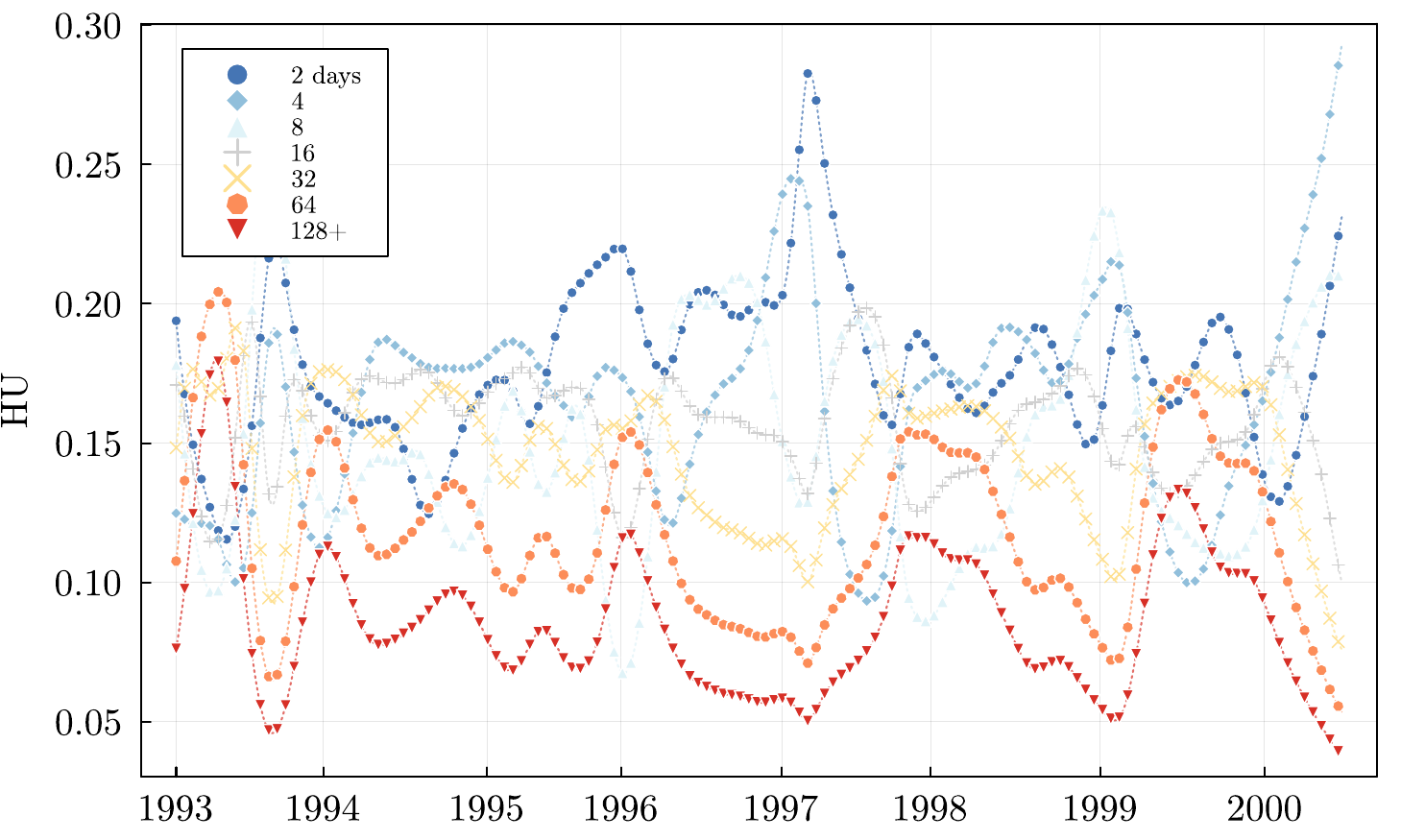}
            \end{center}
             \caption{\footnotesize{Left: Realized volatilities for crude oil (CL), natural gas (NG) and gasoline (HU). Right: Time-varying persistence structure of CL (top), NG (middle), HU (bottom). The plot shows the ratios of $\widehat{\beta}^{j}(t/T,1)/\sum_j \widehat{\beta}^{j}(t/T,1)$ on the y-axis, with $j$ corresponding to 2, 4, 8, 16, 32, 64, 128+ days persistence of the shocks represented by cold to warm colours, respectively, over the period from January 1993 to December 1999 on the x-axis.}}
        \label{fig:persistence_structure_period1}
\end{figure}

While 2010-2022 period was characterised by more persistent shocks driving energy volatility, together with turbulent times and high volumes, we would like to challenge our model with the calm period of the 1990s, when energy futures enjoyed much less volatile prices and much lower volumes. The trades were recorded with a second time stamp during a regular hour when the floor was open. Note that we use unleaded gasoline (HU) futures instead of RBOB gasoline futures, which were introduced after December 2006 only. During this period, we also exclude trades executed on US federal holidays, 24-26 December and 31 December to 2 January, due to low liquidity on these days, which could lead to an estimation bias. Data cover the period from January 1993 to December 1999. The left column of the figure (\ref{fig:persistence_structure_period1}) shows the time series for all three commodities, and it is immediately apparent that the series are much more stable and less volatile.

Similarly to the previous period, we use the multiscale impulse response functions from the TV-EWD model to capture the evolution of the persistence structure of volatility and the persistence of the shocks that generate the volatility time series over this period. Similarly to the previous section, we report the persistence at a given scale $j$ as the persistence share of that scale relative to the persistence at other scales. 

Figure (\ref{fig:persistence_structure_period1}) shows a stark contrast in the persistence structures compared to the previous turbulent period 2010-2022 discussed in the previous section. The main change is that components with higher persistence play a weaker role in the time series and, conversely, the series seem to be driven by transitory shocks. This is important information for modelling purposes.

\subsubsection{Volatility forecasts}

The quieter period 1993-1999 is mainly driven by transitory shocks, but at the same time we can observe some change in the structure. Therefore, our TV-EWD model, which is able to capture the non-trivial dynamics, can also be useful for forecasting, although with increasing complexity it may lose its forecasting power, especially because the model is useful for data with strong long-term persistence.

To compare the forecasting performance of the TV-EWD model, we again use the same models in the same setting, saving the first 700 observations for the in-sample fit and leaving the rest for the out-of-sample forecast period.

Note that the kernel width is set to 0.3 and the depth of decomposition to 7 scales, which is similar for all TV-EWD model settings. This means that the longest horizon considered is 128 and more days (128+). However, the TV-AR(p) model differs for both forecast horizons and commodities. The TV-AR(p) model for CL has $p=3$ for all forecasts. For NG, the TV-AR(p) model has $p=5$ for the 1- and 5-day-ahead forecasts and $p=3$ for the 22-day-ahead forecast, and for HU the TV-AR(p) model has $p=2$ for the 1-day-ahead forecast and $p=5$ for the 5- and 22-day-ahead forecasts.

Table \ref{tab:RV_1} summarises the results of volatility forecasting performance over this period. Again, the TV-HAR model is used as the benchmark and all reported losses are relative to the TV-HAR. In general, the TV-EWD model's ability to outperform all alternatives is weaker during this calm period compared to the turbulent period of 2010-2022. We attribute this mainly to the fact that the high persistence component is weak and therefore time series driven by transitory shocks are more difficult to forecast over multiple horizons. TV-EWD provides improved forecasts in all cases, but the forecast errors are not always significantly different from other models.

\begin{table}[ht!]
\footnotesize
\caption{\footnotesize{The 1993-1999 period: Root mean square error (RMSE) and mean absolute error (MAE) of the TV-EWD model compared to \cite{ortu2020}'s extended world decomposition (EWD), time-varying autoregression (TV-AR(3)). All errors are relative to the time-varying HAR (TV-HAR) model of \cite{corsi2009} over $h=1$, $h=5$ and $h=22$. Thus, the value bellow one implies that the corresponding model outperforms the TV-HAR benchmark. We use $*$,$**$ and $***$ to denote that a competing model has significantly lower MSE and MAE (is better) in comparison to the TV-HAR model at 90\%, 95\% and 99\% significance levels, respectively. We use $\dag$,$\dag \dag$ and $\dag \dag \dag$ to denote that a competing model has significantly higher MSE and MAE (is worse) in comparison to the TV-HAR model at 90\%, 95\% and 99\% significance levels, respectively.
}}

\begin{center}
\begin{tabular}{lllllllll}
\toprule
& & \multicolumn{3}{c}{RMSE} & & \multicolumn{3}{c}{MAE} \\
\cmidrule(r){3-5} \cmidrule(r){7-9}
& & $h=1$ & $h=5$ & $h=22$ & & $h=1$ & $h=5$ & $h=22$ \\
\cmidrule(r){3-5} \cmidrule(r){7-9}
& EWD & 1.122$^{\dag \dag \dag}$ & 1.206$^{\dag \dag \dag}$ & 1.231$^{\dag \dag \dag}$ & & 1.128$^{\dag \dag \dag}$ & 1.301$^{\dag \dag \dag}$ & 1.354$^{\dag \dag \dag}$ \\
CL & TV-AR3 & 1.013 & 1.029 & 0.972 & & 1.024$^{\dag \dag}$ & 1.062$^{\dag \dag \dag}$ & 1.013\\
& TV-EWD & 0.995 & 0.975$^{*}$ & 0.978 & & 0.990 & 0.998 & 1.002\\
\cmidrule(r){3-5} \cmidrule(r){7-9}

& EWD & 1.055$^{\dag \dag \dag}$ & 1.005 & 0.946$^{**}$ & & 1.093$^{\dag \dag \dag}$ & 1.056$^{\dag \dag}$ & 0.914$^{***}$ \\
NG & TV-AR3 & 1.032$^{\dag \dag \dag}$ & 1.081$^{\dag \dag \dag}$ & 1.065$^{\dag \dag \dag}$ & & 1.050$^{\dag \dag \dag}$ & 1.158$^{\dag \dag \dag}$ & 1.063$^{\dag \dag \dag}$\\
& TV-EWD & 1.002 & 0.968 & 0.937$^{***}$ & & 1.003 & 0.995 & 0.960$^{**}$\\
\cmidrule(r){3-5} \cmidrule(r){7-9}

& EWD & 1.080$^{\dag \dag \dag}$ & 1.128$^{\dag \dag \dag}$ & 1.247$^{\dag \dag \dag}$ & & 1.106$^{\dag \dag \dag}$ & 1.181$^{\dag \dag \dag}$ & 1.351$^{\dag \dag \dag}$\\
HU & TV-AR3 & 1.014 & 1.042 & 1.047 & & 1.025$^{\dag \dag}$ & 1.093$^{\dag \dag \dag}$ & 1.134$^{\dag \dag \dag}$\\
& TV-EWD & 0.998 & 0.964$^{*}$ & 0.918$^{*}$ & & 0.998 & 0.979$^{*}$ & 0.976\\
\bottomrule
\end{tabular}
\end{center}
\label{tab:RV_1}
\end{table}

\section{Conclusion}
 \label{sec:conclusion}

In this paper, we study volatility of major energy commodities and identify shocks with heterogeneous persistence that vary smoothly over time. Based on such dynamics, we construct a forecasting model that significantly outperforms alternative models, especially at longer horizons. The model provides valuable information about the fundamental behaviour of the volatility time series and is useful for both practitioners and policy makers.

As the paper introduces a novel decomposition of the volatility time series, it also opens up new avenues for further research. It would be interesting to learn about the dynamic persistence structure of other important energy data and, in particular, to use them in other forecasting or economic models.

Our results can be useful to market participants, financial analysts and policymakers trying to understand energy economics data in several ways, as our approach opens up new avenues for modelling and forecasting in the energy economics literature. Researchers can use our model to understand the dynamically heterogeneous structure of shocks driving different time series, and then adjust the models according to the data. Especially interesting venue of the research is linking the identified components in the data with economic models. One can also use our model to identify transitory and persistent components in the data. Finally, one can use the model to improve forecasts in a number of problems where the data contain rich persistence structures. It would be interesting to use our approach in combination with machine learning.

\bibliography{ext_Wold}

\begin{thebibliography}{}

\bibitem[\protect\citeauthoryear{Andersen, Bollerslev, Diebold, and
  Labys}{Andersen et~al.}{2003}]{andersen2003modeling}
Andersen, T.~G., T.~Bollerslev, F.~X. Diebold, and P.~Labys (2003).
\newblock Modeling and forecasting realized volatility.
\newblock {\em Econometrica\/}~{\em 71\/}(2), 579--625.

\bibitem[\protect\citeauthoryear{Arouri, Lahiani, L{\'e}vy, and Nguyen}{Arouri
  et~al.}{2012}]{arouri2012forecasting}
Arouri, M. E.~H., A.~Lahiani, A.~L{\'e}vy, and D.~K. Nguyen (2012).
\newblock Forecasting the conditional volatility of oil spot and futures prices
  with structural breaks and long memory models.
\newblock {\em Energy Economics\/}~{\em 34\/}(1), 283--293.

\bibitem[\protect\citeauthoryear{Baillie, Chung, and Tieslau}{Baillie
  et~al.}{1996}]{baillie1996analysing}
Baillie, R.~T., C.-F. Chung, and M.~A. Tieslau (1996).
\newblock Analysing inflation by the fractionally integrated arfima--garch
  model.
\newblock {\em Journal of applied econometrics\/}~{\em 11\/}(1), 23--40.

\bibitem[\protect\citeauthoryear{Bandi, Chaudhuri, Lo, and Tamoni}{Bandi
  et~al.}{2021}]{bandi2021}
Bandi, F.~M., S.~E. Chaudhuri, A.~W. Lo, and A.~Tamoni (2021).
\newblock Spectral factor models.
\newblock {\em Journal of Financial Economics\/}~{\em 142\/}(1), 214--238.

\bibitem[\protect\citeauthoryear{Bandi and Tamoni}{Bandi and
  Tamoni}{2022}]{bandi2022spectral}
Bandi, F.~M. and A.~Tamoni (2022).
\newblock Spectral financial econometrics.
\newblock {\em Econometric Theory\/}~{\em 38\/}(6), 1175--1220.

\bibitem[\protect\citeauthoryear{Barunik and Vacha}{Barunik and
  Vacha}{2023}]{barunik2023persistence}
Barunik, J. and L.~Vacha (2023).
\newblock The dynamic persistence of economic shocks.
\newblock {\em Available at SSRN 4467110\/}.

\bibitem[\protect\citeauthoryear{Bollerslev and Engle}{Bollerslev and
  Engle}{1993}]{bollerslev1993common}
Bollerslev, T. and R.~F. Engle (1993).
\newblock Common persistence in conditional variances.
\newblock {\em Econometrica: Journal of the Econometric Society\/}~{\em
  61\/}(1), 167--186.

\bibitem[\protect\citeauthoryear{Charfeddine}{Charfeddine}{2014}]{charfeddine2014true}
Charfeddine, L. (2014).
\newblock True or spurious long memory in volatility: Further evidence on the
  energy futures markets.
\newblock {\em Energy policy\/}~{\em 71}, 76--93.

\bibitem[\protect\citeauthoryear{Corsi}{Corsi}{2009}]{corsi2009}
Corsi, F. (2009).
\newblock A simple approximate long-memory model of realized volatility.
\newblock {\em Journal of Financial Econometrics\/}~{\em 7\/}(2), 174--196.

\bibitem[\protect\citeauthoryear{Dahlhaus}{Dahlhaus}{1996}]{dahlhaus1996}
Dahlhaus, R. (1996).
\newblock On the kullback-leibler information divergence of locally stationary
  processes.
\newblock {\em Stochastic processes and their applications\/}~{\em 62\/}(1),
  139--168.

\bibitem[\protect\citeauthoryear{Elder and Serletis}{Elder and
  Serletis}{2010}]{elder2010oil}
Elder, J. and A.~Serletis (2010).
\newblock Oil price uncertainty.
\newblock {\em Journal of Money, Credit and Banking\/}~{\em 42\/}(6),
  1137--1159.

\bibitem[\protect\citeauthoryear{Fan and Gijbels}{Fan and
  Gijbels}{1996}]{fan1996local}
Fan, J. and I.~Gijbels (1996).
\newblock {\em Local polynomial modelling and its applications: monographs on
  statistics and applied probability 66}, Volume~66.
\newblock CRC Press.

\bibitem[\protect\citeauthoryear{Granger and Ding}{Granger and
  Ding}{1996}]{granger1996varieties}
Granger, C.~W. and Z.~Ding (1996).
\newblock Varieties of long memory models.
\newblock {\em Journal of econometrics\/}~{\em 73\/}(1), 61--77.

\bibitem[\protect\citeauthoryear{Granger and Hyung}{Granger and
  Hyung}{2004}]{granger2004occasional}
Granger, C.~W. and N.~Hyung (2004).
\newblock Occasional structural breaks and long memory with an application to
  the s\&p 500 absolute stock returns.
\newblock {\em Journal of empirical finance\/}~{\em 11\/}(3), 399--421.

\bibitem[\protect\citeauthoryear{Hamilton}{Hamilton}{1983}]{hamilton1983oil}
Hamilton, J.~D. (1983).
\newblock Oil and the macroeconomy since world war ii.
\newblock {\em Journal of political economy\/}~{\em 91\/}(2), 228--248.

\bibitem[\protect\citeauthoryear{Hamilton}{Hamilton}{2020}]{hamilton2020time}
Hamilton, J.~D. (2020).
\newblock {\em Time series analysis}.
\newblock Princeton university press.

\bibitem[\protect\citeauthoryear{Haugom, Langeland, Moln{\'a}r, and
  Westgaard}{Haugom et~al.}{2014}]{haugom2014forecasting}
Haugom, E., H.~Langeland, P.~Moln{\'a}r, and S.~Westgaard (2014).
\newblock Forecasting volatility of the us oil market.
\newblock {\em Journal of Banking \& Finance\/}~{\em 47}, 1--14.

\bibitem[\protect\citeauthoryear{Herrera, Hu, and Pastor}{Herrera
  et~al.}{2018}]{herrera2018forecasting}
Herrera, A.~M., L.~Hu, and D.~Pastor (2018).
\newblock Forecasting crude oil price volatility.
\newblock {\em International Journal of Forecasting\/}~{\em 34\/}(4), 622--635.

\bibitem[\protect\citeauthoryear{Kang and Yoon}{Kang and
  Yoon}{2013}]{kang2013modeling}
Kang, S.~H. and S.-M. Yoon (2013).
\newblock Modeling and forecasting the volatility of petroleum futures prices.
\newblock {\em Energy Economics\/}~{\em 36}, 354--362.

\bibitem[\protect\citeauthoryear{Kilian}{Kilian}{2009}]{kilian2009not}
Kilian, L. (2009).
\newblock Not all oil price shocks are alike: Disentangling demand and supply
  shocks in the crude oil market.
\newblock {\em American Economic Review\/}~{\em 99\/}(3), 1053--1069.

\bibitem[\protect\citeauthoryear{Le, Boubaker, Bui, and Park}{Le
  et~al.}{2023}]{le2023volatility}
Le, T.-H., S.~Boubaker, M.~T. Bui, and D.~Park (2023).
\newblock On the volatility of wti crude oil prices: A time-varying approach
  with stochastic volatility.
\newblock {\em Energy Economics\/}~{\em 117}, 106474.

\bibitem[\protect\citeauthoryear{Lu, Ma, Li, and Huang}{Lu
  et~al.}{2022}]{LU2022102218}
Lu, F., F.~Ma, P.~Li, and D.~Huang (2022).
\newblock Natural gas volatility predictability in a data-rich world.
\newblock {\em International Review of Financial Analysis\/}~{\em 83}, 102218.

\bibitem[\protect\citeauthoryear{Ma, Wahab, Huang, and Xu}{Ma
  et~al.}{2017}]{ma2017forecasting}
Ma, F., M.~I.~M. Wahab, D.~Huang, and W.~Xu (2017).
\newblock Forecasting the realized volatility of the oil futures market: A
  regime switching approach.
\newblock {\em Energy Economics\/}~{\em 67}, 136--145.

\bibitem[\protect\citeauthoryear{Ortu, Severino, Tamoni, and Tebaldi}{Ortu
  et~al.}{2020}]{ortu2020}
Ortu, F., F.~Severino, A.~Tamoni, and C.~Tebaldi (2020).
\newblock A persistence-based wold-type decomposition for stationary time
  series.
\newblock {\em Quantitative Economics\/}~{\em 11\/}(1), 203--230.

\bibitem[\protect\citeauthoryear{Ozdemir, Gokmenoglu, and Ekinci}{Ozdemir
  et~al.}{2013}]{ozdemir2013persistence}
Ozdemir, Z.~A., K.~Gokmenoglu, and C.~Ekinci (2013).
\newblock Persistence in crude oil spot and futures prices.
\newblock {\em Energy\/}~{\em 59}, 29--37.

\bibitem[\protect\citeauthoryear{S{\'e}vi}{S{\'e}vi}{2014}]{sevi2014forecasting}
S{\'e}vi, B. (2014).
\newblock Forecasting the volatility of crude oil futures using intraday data.
\newblock {\em European Journal of Operational Research\/}~{\em 235\/}(3),
  643--659.

\bibitem[\protect\citeauthoryear{St{\u{a}}ric{\u{a}} and
  Granger}{St{\u{a}}ric{\u{a}} and
  Granger}{2005}]{stuaricua2005nonstationarities}
St{\u{a}}ric{\u{a}}, C. and C.~Granger (2005).
\newblock Nonstationarities in stock returns.
\newblock {\em Review of economics and statistics\/}~{\em 87\/}(3), 503--522.

\bibitem[\protect\citeauthoryear{Wang and Wu}{Wang and Wu}{2012}]{wang2012long}
Wang, Y. and C.~Wu (2012).
\newblock Long memory in energy futures markets: Further evidence.
\newblock {\em Resources Policy\/}~{\em 37\/}(3), 261--272.

\bibitem[\protect\citeauthoryear{Wang, Wu, and Yang}{Wang
  et~al.}{2016}]{wang2016forecasting}
Wang, Y., C.~Wu, and L.~Yang (2016).
\newblock Forecasting crude oil market volatility: A markov switching
  multifractal volatility approach.
\newblock {\em International Journal of Forecasting\/}~{\em 32\/}(1), 1--9.

\bibitem[\protect\citeauthoryear{Wen, Gong, and Cai}{Wen
  et~al.}{2016}]{wen2016forecasting}
Wen, F., X.~Gong, and S.~Cai (2016).
\newblock Forecasting the volatility of crude oil futures using har-type models
  with structural breaks.
\newblock {\em Energy Economics\/}~{\em 59}, 400--413.

\bibitem[\protect\citeauthoryear{Wold}{Wold}{1938}]{wold1938}
Wold, H. (1938).
\newblock {\em A study in the analysis of stationary time series}.
\newblock Almquist \& Wiksells Boktryckeri.

\bibitem[\protect\citeauthoryear{Zhang, Wei, Zhang, and Jin}{Zhang
  et~al.}{2019}]{zhang2019forecasting}
Zhang, Y., Y.~Wei, Y.~Zhang, and D.~Jin (2019).
\newblock Forecasting oil price volatility: Forecast combination versus
  shrinkage method.
\newblock {\em Energy Economics\/}~{\em 80}, 423--433.

\end{thebibliography}
\bibliographystyle{chicago}


\end{document}